\documentclass[aps,prd,groupedaddress,showpacs,nofootinbib,amssymb,balancelastpage,preprintnumbers,floatfix]{revtex4}

\usepackage[dvipdfmx]{graphicx}
\usepackage{graphicx,bm,color}

\usepackage{amsmath}
\usepackage{amssymb}
\usepackage{amsfonts}
\usepackage{cases}
\usepackage{cancel}
\usepackage[normalem]{ulem}
\usepackage{subfigure}
\usepackage{here}
\usepackage{color}
\usepackage{comment}
\usepackage{hyperref}

\usepackage{tikz}
\usetikzlibrary{matrix}

\allowdisplaybreaks[3]

\newcommand{\cred}[1]{{\color{black}#1}}

\begin{document}
	
%\preprint{APS/123-QED}

\title{Implications of electromagnetic scale anomaly to QCD chiral phase transition in smaller quark mass regime: $T_{\mathrm{pc}}$ does not drop with eB }

\author{Yuanyuan Wang}\thanks{{\tt yuanyuanw23@jlu.edu.cn}}
\affiliation{Center for Theoretical Physics and College of Physics, Jilin University, Changchun, 130012,
China}

\author{Mamiya Kawaguchi}\thanks{{\tt mamiya@aust.edu.cn}} 
      \affiliation{ 
Center for Fundamental Physics, School of Mechanics and Physics,
Anhui University of Science and Technology, Huainan, Anhui 232001, People’s Republic of China
}

\author{Shinya Matsuzaki}\thanks{{\tt synya@jlu.edu.cn}}
\affiliation{Center for Theoretical Physics and College of Physics, Jilin University, Changchun, 130012, China}%

\author{Akio Tomiya}\thanks{{\tt akio@yukawa.kyoto-u.ac.jp}} 
\affiliation{Department of Information and Mathematical Sciences, Tokyo Woman’s Christian  University, Tokyo 167-8585, Japan} 
\affiliation{RIKEN Center for Computational Science, Kobe 650-0047, Japan} 
%RIKEN BNL Research center, Brookhaven National Laboratory, Upton, NY, 11973, USA 

\begin{abstract}

The decrease of the chiral pseudocritical temperature $T_{\mathrm{pc}}$ with an applied strong magnetic field has been extensively investigated by various QCD low-energy effective models and lattice QCD at physical point. 
We find that this decreasing feature may not hold in the case with a weak magnetic field and still depends on quark masses: when the quark masses get smaller, $T_{\mathrm{pc}}$ turns to increase with the weak magnetic field. This happens due to the significant electromagnetic-scale anomaly contribution in the thermomagnetic medium. 
We demonstrate this salient feature by employing the Nambu-Jona-Lasinio model with 
2 + 1 quark flavors including the electromagnetic-scale anomaly contribution. 
We observe that at $(m_{0c}, m_{sc}) \simeq (2, 20) \mathrm{MeV}$ for the isospin symmetric mass for up and down quarks, $m_0$, and the strange quark mass, $m_s$, $T_{\mathrm{pc}}$ decreases with the magnetic field if the quark masses exceed the critical values, and increases as the quark masses become smaller. 
Related cosmological implications, arising when the supercooled electroweak phase transition or dark QCD cosmological phase transition is considered along with a primordial magnetic field, are also briefly addressed.

\end{abstract}

\maketitle

\section{Introduction} 

The chiral phase transition in QCD with a possible presence of magnetic fields has extensively been investigated in light of deeply understanding heavy ion collision experiments. 
External strong magnetic fields make significant effects on the hot QCD plasma and 
the nature of the chiral phase transition, such as what are called 
the reduction of the chiral pseudocritical temperature and the inverse magnetic catalysis~\cite{Bali:2011qj,Bornyakov:2013eya, Bali:2014kia, Tomiya:2019nym, DElia:2018xwo, Endrodi:2019zrl}, and so forth.

Also in the early hot Universe, the strong magnetic field could be generated via cosmological first-order phase transitions~\cite{Vachaspati:1991nm,Enqvist:1993np,Grasso:1997nx,Grasso:2000wj,Ellis:2019tjf,Zhang:2019vsb,Di:2020kbw,Yang:2021uid}, 
or inflationary scenarios~\cite{Turner:1987vd,Garretson:1992vt,Anber:2006xt,Domcke:2019mnd,Domcke:2019qmm,Patel:2019isj,Domcke:2020zez,Shtanov:2020gjp,Okano:2020uyr,Cado:2021bia,Kushwaha:2021csq,Gorbar:2021rlt,Gorbar:2021zlr,Gorbar:2021ajq,Fujita:2022fwc}, which would be redshifted to a later epoch when the Universe undergoes the QCD phase transition. 
The QCD phase transition epoch may thus be magnetized by the redshifed primordial magnetic field, and the chiral phase transition could take place in the thermal bath with a background magnetic field  
coupled to the quarks. 
The strength of such a cosmic magnetic field background might be as small as or smaller than the intrinsic scale of QCD $\Lambda_{\rm QCD} 
\sim (200 - 400)$ MeV. 
In that case,  
the QCD phase transition cannot simply be dominated by physics of  
the lowest Landau levels longitudinally polarized along the direction 
parallel to the magnetic field, which leads to the dimensional reduction into 2 + 2 system  
and can support interpretations on the 
inverse magnetic catalysis (see, e.g., ~\cite{Bandyopadhyay:2020zte}).

Thermal QCD with such a weak magnetic field has not yet been well explored. 
Lattice QCD has observed the inverse magnetic catalysis and the decrease of $T_{\rm pc}$ as the magnetic field strength $eB$ gets larger  
at high enough $T = {\cal O}(100\,{\rm MeV})$~\cite{Bali:2011qj,Bornyakov:2013eya, Bali:2014kia, Tomiya:2019nym, DElia:2018xwo, Endrodi:2019zrl}. 
On lattices, the size of $eB$ is bounded from below as $eB \gtrsim T^2$ due to the currently 
applied method to create the magnetic field~\cite{Bali:2011qj}, 
which works only for the physical pion mass simulations at finite temperatures. 
Thus applying a weak $eB$ at high $T$ with much smaller pion masses, particularly close to the chiral limit, is challenging due to the higher numerical cost, 
and it is still uncertain whether the normal or inverse magnetic catalysis persists and $T_{\rm pc}$ gets increased or reduced with $eB$.

Recently it has been clarified, based on the linear sigma model and Nambu-Jona-Lasinio (NJL) model with the mean field approximation (MFA), 
that the electromagnetic-scale anomaly effect is not negligible in a weak magnetic field and is shown to have important implications for the chiral phase transition properties regarding the quark mass dependence~\cite{Kawaguchi:2021nsa,YuanyuanWang:2022nds}. 
Without applying an external magnetic field, 
in the massless-two flavor case those chiral effective models (in the MFA) predict  
the chiral phase transition of second order, and in the massless three-flavor case, 
it goes like the first order type~\footnote{
Regarding the current status on the lattice QCD study with $eB=0$, 
in the case with very light three flavors,   
the first-order scenario has not yet been excluded~\cite{Dini:2021hug,Bazavov:2017xul} for 
highly improved staggered fermions. 
Another analysis with unimproved staggered fermions 
suggests second-order like phase transitions for $N_f \le 6$~\cite{Cuteri:2021ikv}.  
 Thus the first-order nature in the case with $N_f=3$ has not been conclusive yet even using lattice QCD.}  
The trend drastically gets altered, however, in the presence of a weak magnetic field 
(with $\sqrt{eB} \lesssim \Lambda_{\rm QCD} \sim (200 - 400)$ MeV): 
when $eB$ exceeds a critical strength (depending on those models), 
the chiral phase transition turns to be crossover-like in the massless-two flavor case~\cite{Kawaguchi:2021nsa}, and the first order nature is gone to be of the second order or crossover-like also in the massless-three flavor case ~\cite{YuanyuanWang:2022nds}.

The disappearance of the first-order feature 
is tied with the crucial presence 
of the electromagnetic scale anomaly contributing to the chiral symmetry breaking at around the criticality.  
As has been emphasized in the literature~\cite{YuanyuanWang:2022nds}, 
this trend is irrespective to details of the parameter setting of the chiral effective model, hence 
is fairly generic and can be applied to a wide class of QCD-like theories,  
which might still hold even beyond the mean-field approximation in the NJL description:  
for instance, 
QCD can be replaced with dark QCD and the electromagnetic field with a dark photon field which is typically the portal between the dark and the Standard Model sectors. 
The disappearance of the first order nature then makes crucial impacts on modeling 
the dark QCD scenario in light of the gravitational wave production sourced from 
the supercooled cosmological phase transition, such as those discussed in the literature~\cite{Holthausen:2013ota,Ametani:2015jla,Aoki:2017aws,Bai:2018dxf,Helmboldt:2019pan,Reichert:2021cvs,Archer-Smith:2019gzq,Aoki:2019mlt,Dvali:2019ewm,Easa:2022vcw,Tsumura:2017knk},  
as well as would help deeply understanding the chiral phase structure in ordinary QCD. 
Furthermore, QCD in the Standard Model with (almost) massless quarks can be viable also in 
the supercooled electroweak phase transition scenarios, in which QCD first triggers the electroweak 
symmetry breaking via the chiral phase transition along with massless six quarks~\cite{Witten:1980ez,Iso:2017uuu,Hambye:2018qjv,Sagunski:2023ynd}.

Given such cosmological interests also at hand,  
in this paper, we extend the analysis in~\cite{YuanyuanWang:2022nds} 
by investigating the weak $eB$ dependence on $T_{\mathrm{pc}}$ with the quark masses varied. 
We find that $T_{\rm pc}$ may not decrease with the magnetic field in the smaller quark mass regime.
This happens due to the presence of the magnetic catalysis with a weak $e B$
and the chiral first-order phase transition in the smaller quark mass regime. As quarks get heavier to go off the first order regime, $T_{\mathrm{pc}}$ turns to decrease with $e B$. 
We demonstrate this salient feature by employing 
a NJL model with the lightest three flavor quarks 
in the MFA with electromagnetic scale anomaly.

Conventionally, the reduction of $T_{\mathrm{pc}}$ with respect to the magnetic field 
is thought to be identical to 
the inverse magnetic catalysis. However, the literature~\cite{DElia:2018xwo,Ali:2024mnn} suggests that at a large enough pion mass the inverse magnetic catalysis, i.e., the decrease of the quark condensate with the increase of the  magnetic field will disappear, whereas the reduction of $T_{\mathrm{pc}}$ with the increase of the magnetic field will still persist. 
This might imply the universal feature of the reduction of $T_{\rm pc}$ in the quark mass plane, when projected onto the conventional Columbia plot~\cite{Brown:1990ev}.  
However, what we argue in the present paper is to propose a counter case to this seemingly universal reduction feature, 
in a sense that the reduction of $T_{\rm pc}$ does not drop with the increase of $eB$ 
if quark masses get small enough. 
We observe that at $(m_{0c}, m_{sc}) \simeq (2, 20) \, \mathrm{MeV}$ for the isospin symmetric up and down quark mass $m_0$ and the strange quark mass $m_s$, $T_{\mathrm{pc}}$ decreases with the magnetic field if the quark masses are greater than those values, and increases with the magnetic field if the quark masses are less than them.

Our finding helps deeply understand the chiral phase transition in the wider 
parameter space in QCD, including quark masses, temperature, and also the magnetic field strength. 
The increase of $T_{\rm pc}$ could also help investigate the cosmological consequences from the thermomagnetized QCD or dark QCD in the thermal history of the universe, as briefly noted in the Summary and Discussion section.

\section{NJL model with electromagnetic scale anomaly}\label{Sec-II}

We start with introducing the NJL model with the 2 + 1 flavors, which is also  
coupled to a constant external magnetic field. 
The Lagrangian describing the model for the quark triplet $\psi=(u, d, s)^T$ 
is written as follows:
\begin{equation}
\begin{aligned}
\mathcal{L}= & \bar{\psi}\left(i \gamma^\mu D_\mu-\mathbf{m}\right) \psi+\sum_{a=0}^8 G\left\{\left(\bar{\psi} \lambda^a \psi\right)^2+\left(\bar{\psi} i \gamma^5 \lambda^a \psi\right)^2\right\} \\
& -K\left[\operatorname{det} \bar{\psi}\left(1+\gamma_5\right) \psi+\operatorname{det} \bar{\psi}\left(1-\gamma_5\right) \psi\right] ,
\end{aligned}
\label{Lag:NJL}
\end{equation}
where the current quark mass matrix ${\bf m}$ takes the form $\mathbf{m}=\operatorname{diag}\left\{m_0, m_0, m_s\right\}$, and $\lambda^a(a=0, \ldots .8)$ represents the 
Gell-Mann matrices in the flavor space with $\lambda^0=\sqrt{2 / 3}\operatorname{diag}(1,1,1)$. 
The covariant derivative is defined as $D_\mu=\partial_\mu-i q A^{\rm em}_\mu$, which includes the external electromagnetic field $A^{\rm em}_\mu$, with the electromagnetic charge matrix \cred{$q={\rm diag}\{q_u, q_d, q_s \} = e \cdot {\rm diag}\{2 /3,-1 /3 - 1/3 \}$}.  
The constant magnetic field $B$ is embedded in $A_\mu^{\rm em}$ as $A^{\rm em}_\mu=(0,0, B x, 0)$. 
In Eq.(\ref{Lag:NJL}), the four-fermion interaction term with the coupling $G$ 
keeps the full chiral $U(3)_L \times U(3)_R$ symmetry, while 
the Kobayashi-Maskawa-’t Hooft (KMT) determinant term $\propto K$~\cite{Kobayashi:1970ji,Kobayashi:1971qz,tHooft:1976snw}, which arises from the QCD instanton effect coupled to quarks. is invariant under the $SU(3)_L \times SU(3)_R \times U(1)_V$ symmetry, but is anomalous for the $U(1)_A$ symmetry.

We introduce the three-dimensional momentum cutoff $\Lambda$ to regularize the NJL model in the MFA. 
We refer to empirical hadron observables in the isospin symmetric limit 
at $e B=T=0$~\cite{Rehberg:1995kh}:  
the pion mass $m_\pi \simeq 135 \,   \mathrm{MeV}$, the pion decay constant $f_\pi=92.4 \, \mathrm{MeV}$, the $\eta^{\prime}$ mass $m_{\eta^{\prime}} \simeq 957.8 \, \mathrm{MeV}$, and the kaon mass $m_K \simeq 497.7 \, \mathrm{MeV}$. 
This together with $m_0=5.5 \, \mathrm{MeV}$ as inputs fixes 
the model parameters  
as $\Lambda=0.6023 \mathrm{GeV}$, $G \Lambda^2=1.835$, and $K \Lambda^5=12.36$. 
Then we have $m_s\simeq 140.7\,  \mathrm{MeV}$ and the dynamical mass for up an down quarks,  
$m_{\rm dyn} \simeq 367.7$ MeV. 
In the present study, we shall call this parameter set with $(m_0=5.5 \, \mathrm{MeV}, m_s = 140.7 \, \mathrm{MeV}$) the physical point in QCD.

Specifically important to note is that the thermal chiral phase transition in the NJL can be controlled by 
the following set of the parameter ratios: 
\begin{align} 
	G\Lambda^2 \simeq 1.835, \qquad K\Lambda^5 \simeq 12.36
\,, \qquad 
	\Lambda/f_\pi \simeq  6.518   	
	\,. \label{parameter-set}
	\end{align} 
This is what is called the QCD-monitored parameter setup,  
%in Eqs.(\ref{parameters1}) and  (\ref{parameters2}) 
which forms a wide class of QCD-like theories in the NJL-like description given by scaling up of ordinary QCD with respect to $\Lambda$ or $f_\pi$~\cite{YuanyuanWang:2022nds}. 
%A similar parameter setup derived based on a different regularization scheme
%has been applied in dark QCD scenarios with use of the scaling up
A similar parameter setting obtained based on a different regularization scheme has been employed for dark QCD scenarios using scaling up~\cite{Holthausen:2013ota,Ametani:2015jla,Aoki:2017aws,Reichert:2021cvs,Helmboldt:2019pan,Aoki:2019mlt}.  
Note also that as has been emphasized in~\cite{YuanyuanWang:2022nds}, 
%the present analysis is actually irrespective to 
%the size of $e$ itself, because the external-magnetic field strength 
%contributes to the chiral phase transition with keeping the gauge-invariant form $F_{12} = eB$ 
%without separation between $e$ and $B$. 
the present analysis is actually irrespective to the size of $e$ itself, as the strength of the external magnetic field contributes to the chiral phase transition while preserving the gauge-invariant form $F_{12} = eB$  without separating $e$ and $B$.

To evaluate the thermal and magnetic corrections to the chiral phase transition, we employ the imaginary time formalism and the Landau level decomposition with the following replacements:
\begin{equation}
\begin{aligned}
p_0 & \qquad \leftrightarrow \qquad i \omega_N =i(2 N +1) \pi T, \\
\int \frac{d^4 p}{(2 \pi)^4} & \qquad \leftrightarrow  \qquad i T \sum_{N=-\infty}^{\infty} \int \frac{d^3 p}{(2 \pi)^3} \\
& \qquad \leftrightarrow \qquad i T \sum_{N=-\infty}^{\infty} \sum_{n=0}^{\infty}\alpha_n \frac{\left|q_f  B\right|}{4 \pi} \int_{-\infty}^{\infty} \frac{d p_z}{2 \pi} f_{\Lambda}(p_{z},n).
\end{aligned}
\label{replacement} 
\end{equation}
$\omega_N$ (with $N$ being integers) denotes the Matsubara frequency; 
$n$ represents the Landau levels (which will be summed up to 500 in the numerical analysis); 
$\alpha_n=2-\delta_{n, 0}$ corresponds to the spin degeneracy factor with respect to the Landau levels. 
With those replacements, 
it is straightforward to extend the chiral phase transition in the vacuum to the case with $T\neq 0$ and $eB \neq 0$. 
In the present study, as seen from Eq.(\ref{replacement}), 
we simply apply a conventional soft cutoff scheme~\cite{Frasca:2011zn} to 
regularize the integration over $p_z$ in Eq.(\ref{replacement}), in such a way that 
\begin{equation}
	\begin{aligned}
	f_{\Lambda}(p_{z},n)=\frac{\Lambda^{10}}{\Lambda^{10}+(p^2_{z}+2n\lvert q_{f}B\rvert)^5}
	\,.  
	\end{aligned}
\end{equation}

In addition to the conventional NJL model terms in Eq.(\ref{Lag:NJL}),  
we incorporate the electromagnetic-scale anomaly into the NJL model as in the literature~\cite{Kawaguchi:2021nsa,YuanyuanWang:2022nds}, which takes the form  
\begin{equation}
V_{\mathrm{eff}}^{(\mathrm{Tad})} 
=-\frac{\varphi}{f_{\varphi}} T_\mu^\mu \,,    
\label{Tad}
\end{equation}
Here $\varphi$ denotes the chiral-singlet part of the scalar mesons fields (like {\it a QCD dilaton}),  
associated with the radial component of the nonet-quark bilinear $\bar{q}_f q_{f^{\prime}}$; $T_\mu^\mu$ is the trace of the energy-momentum tensor, which dictates 
the electromagnetic scale anomaly, to be more explicitly evaluated right below; 
$f_\varphi$ stands for the vacuum expectation value of $\varphi$, which will be explicitized below.

In the weak field limit $e B \ll \Lambda_{\rm QCD}^2$, 
the trace anomaly $T_\mu^\mu$ in Eq.(\ref{Tad}) 
takes the form 
\begin{equation}
T_\mu^\mu %\left(M_f, T, e B\right)
=\frac{\beta(e)}{e^3} |e B|^2+\frac{1}{2}N_c \sum_{f=u,d,s}\sum_{n=0}^{\infty} \alpha_n\frac{ q_f^2}{e^2}M^2_f \frac{\left|q_f B\right|}{4 \pi} \int_{-\infty}^{\infty} \frac{d p_z}{2 \pi} F\left(T,e B, M_f\right)|e B|^2. 
\label{tr-anom}
\end{equation}
The first term $\propto \beta(e)$ in Eq.(\ref{tr-anom})  corresponds to the electromagnetic-scale anomaly at the vacuum ($T=eB=0$), which comes in there along with 
the $\beta$ function coefficient of the electromagnetic coupling $e$, as is well known.  
At the one-loop level, the $\beta$ function is given as 
\begin{equation}
\beta(e)=\frac{e}{(4 \pi)^2} \frac{4 N_c}{3} \sum_f q_f^2
\,. 
\end{equation}
 The second term in Eq.(\ref{tr-anom}) is the thermomagnetic part including the Fermi-Dirac thermal distribution function as (for details, see Appendix~\ref{app})~\footnote{
In~\cite{Kawaguchi:2021nsa,YuanyuanWang:2022nds} the thermomagnetic part in the second term of Eq.(\ref{tr-anom}) was evaluated by introducing the typical magnetic field strength $\sqrt{eB}$ as the infrared-loop momentum cutoff, so that the overall weak $eB$ dependence was reduced by one order of magnitude. In the present study, we have refined the evaluation without introduction of such a cutoff, so that the overall $eB$ dependence is of quadratic order as seen in Eq.(\ref{tr-anom}) and the kernel function $F(T,eB,M_f)$ has simply been set by the form of the thermal distribution function as in Eq.(\ref{F}). 
}:  
\begin{equation}
\begin{aligned}
& F\left(T, e B, M_f\right)=\frac{1}{E^5_{f, n}}\cdot\frac{-2}{\exp \left(\frac{E_{f, n}}{T}\right)+1}, \qquad {\rm with} \qquad 
& E^2_{f, n}=p_z^2+2 n\left|q_f B\right|+M_f^2, 
\end{aligned} \label{F} 
\end{equation}
and the constituent (full) quark masses $M_{u,d,s}$ are defined as 
\begin{equation}
\begin{aligned}
M_u & =m_u-4 G \phi_u+2 K \phi_d \phi_s \equiv m_0+\sigma_u, \\
M_d & =m_d-4 G \phi_d+2 K \phi_u \phi_s \equiv m_0+\sigma_d, \\
M_s & =m_s-4 G \phi_s+2 K \phi_u \phi_d \equiv m_s+\sigma_s\,, 
\end{aligned}
\label{M}
\end{equation}
where $\sigma_f$ corresponds to the dynamical quark mass of the $q_f$ quark, which also 
acts as a $\sigma$ (or $\sigma_s$) meson field in the effective potential at the level of the MFA. 
In Eq.(\ref{M}) $\phi_f$ stands for the quark condensate for the $q_f$ quark, 
\cred{which will be evaluated as the solution of the stationary condition for 
the thermomagnetic potential.} 
In terms of $\sigma_f$, the chiral-singlet scalar-meson field $\varphi$ and its vacuum expectation value 
$f_\varphi$ in Eq.(\ref{Tad}) are expressed as 
\begin{align} 
\frac{\varphi}{f_\varphi} = \sqrt{ \frac{\sigma_u^2 + \sigma_d^2 + \sigma_s^2}{\sigma_{u0}^2 + \sigma_{d0}^2 + \sigma_{s0}^2}} 
=
\sqrt{ \frac{\sigma_u^2 + \sigma_d^2 + \sigma_s^2}{2 \sigma_{0}^2  + \sigma_{s0}^2}} 
\,, 
\end{align} 
where $\sigma_{f0}$ represents the vacuum expectation value of $\sigma_f$ at 
$T=eB=0$, and $\sigma_{u0}=\sigma_{d0} = \sigma_{0}$.

When $eB$ gets stronger than the threshold strength, $e B \lesssim \Lambda_{\rm QCD}^2$, 
the electromagnetic scale anomaly tends to be washed out,   
due to the fact that the dynamics of quarks are dominated by 
the lowest Landau level states, which are longitudinally polarized along the magnetic field direction. 
Therefore, the electromagnetic scale anomaly is 
most remarkable in a weak magnetic regime, as first clarified in the literature~\cite{Kawaguchi:2021nsa}.

The constituent quark mass $M_f$ in the second term of Eq.(\ref{tr-anom}) potentially includes higher-order contributions in $eB$, while other possible higher order terms in $eB$ have been 
disregarded in Eq.(\ref{tr-anom}). This may be a crude truncation that we have presently taken. 
In~\cite{Ghosh:2019kmf} a nonperturbative evaluation of the photon polarization function coupled to the chiral-singlet meson-field has been made attempted. It could be possible to make a straightforward computation of the photon polarization in the weak magnetic field. 
The reliability of the present truncation prescription could be evaluated in such a way, 
which is to be left in the future work.

Thus the total thermomagnetic potential in the MFA takes the form
\begin{equation}
\begin{aligned}
    \Omega\left(\phi_u, \phi_d, \phi_s, T, e B\right)=\sum_{f=u, d, s} \Omega_f\left(\phi_f,  T, e B\right)&+2 G\left(\phi_u^2+\phi_d^2+\phi_s^2\right)-4 K \phi_u \phi_d \phi_s\\&+V_{\mathrm{eff}}^{(\mathrm{Tad})}\left(M_f, T, e B\right), 
\end{aligned}
\label{Omega}
\end{equation}
where
\begin{equation}
\begin{aligned}
\Omega_f\left(\phi_f,  T, e B\right) & =- N_c \sum_{n=0}^{\infty} \alpha_n \int_{-\infty}^\infty \frac{d p_z}{2 \pi} \frac{\left|q_f B\right|}{2 \pi} f_{\Lambda}(p_{z},n) 
\Bigg[E_{f,n}  
+2 T \ln \left(1+ e^{-\frac{E_{f,n}}{T}} \right) \Bigg].
\end{aligned}
\end{equation}
The pseudocritical temperature $T_{\rm pc}$ can be determined by the second-order partial derivative of the constituent-up quark mass ($M_u$) with respect to $T$, which is given as 
\begin{equation}
\left.\frac{\partial^2 M_u}{\partial T^2}\right|_{T=T_{\rm pc}}=0 .
\label{def:Tpc}
\end{equation}

\section{Pseudocritical temperature $T_{\rm pc}$} 
In this section, we show that 
due to the presence of the tadpole potential in Eq.(\ref{Tad}), 
$T_{\rm p c}$ does not drop with $e B$ when  quark masses get small enough compared to the physical point values.

\subsection{Analysis at physical point: $T_{\rm pc}$ drops with $eB$}
\label{3.1}

\begin{figure}[t] 
\centering 
\includegraphics[width=0.48\textwidth]{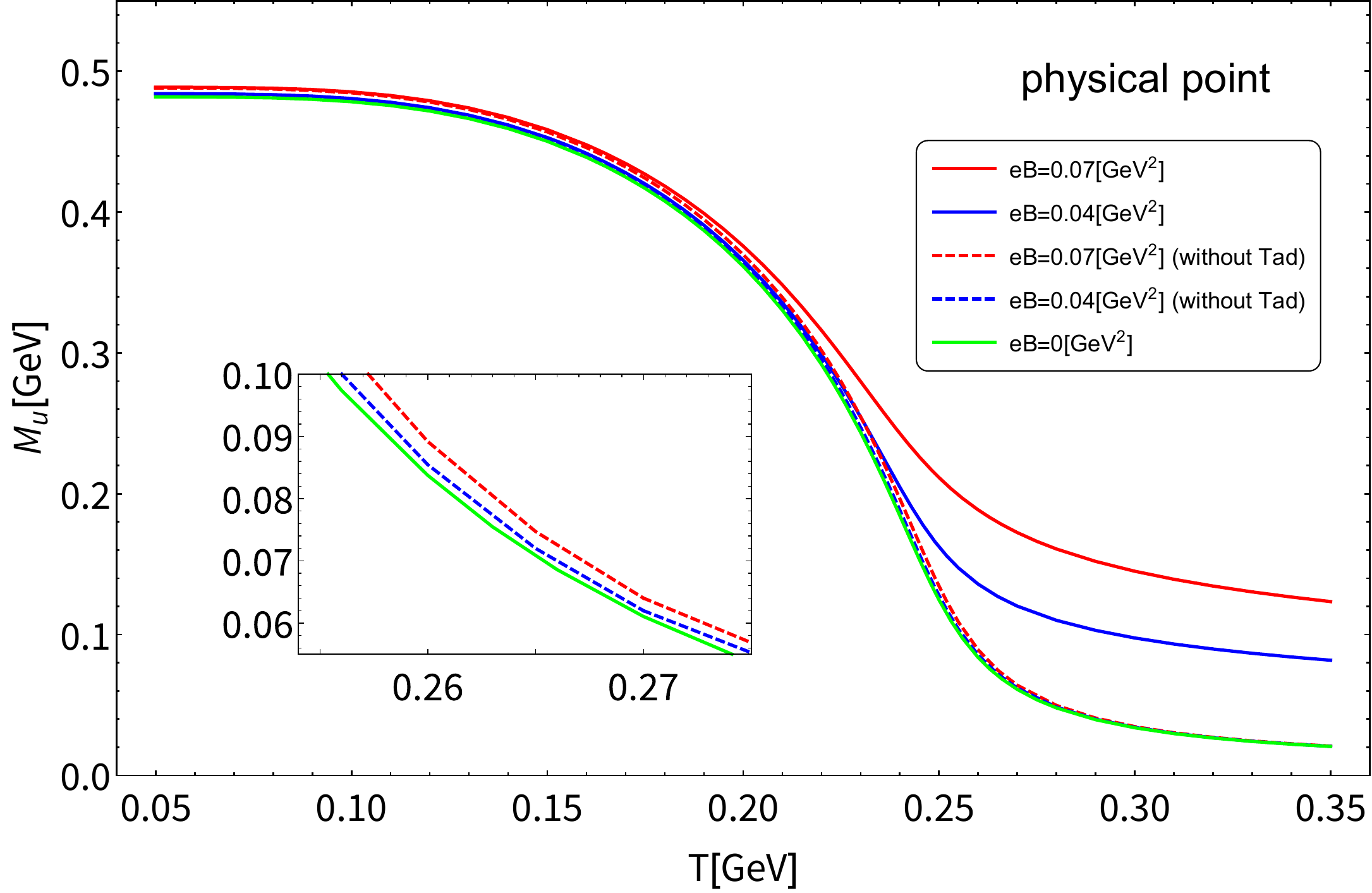}
\includegraphics[width=0.48\textwidth]{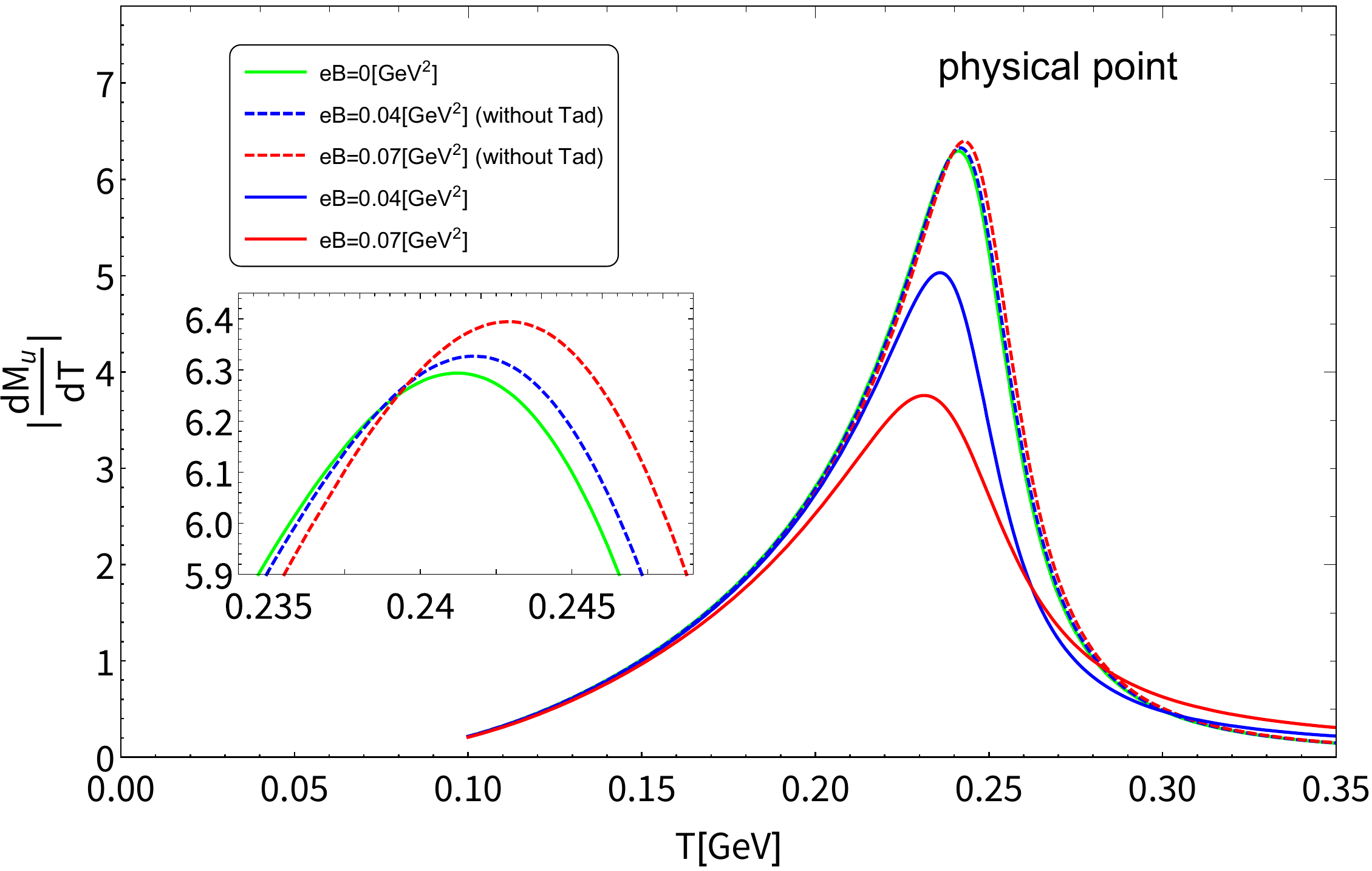}
\caption{Left panel: $T$ dependence of $M_u$ at the physical point with varying $eB$ in the case with (solid curves) or without (dotted curves) the tadpole term in Eq.(\ref{tr-anom}).  
Right panel: the same $eB$ dependence on $\frac{dM_u}{dT}$, in which  
$T_{\rm pc}$ is defined at the extremum as in Eq.(\ref{def:Tpc}). 
} 
\label{M-peak}
\end{figure}

See Fig.~\ref{M-peak}, where we plot $M_u$ and its first derivative with respect to $T$, 
$\frac{dM_u}{dT}$, as a function of $T$, 
with varying $eB$ in the case with or without the tadpole contribution in Eq.(\ref{tr-anom}). 
When the tadpole contribution in Eq.(\ref{tr-anom}) is not taken into account,  
$M_u$ increases slightly with $e B$ in the weak $e B$ range, and $T_{\rm pc}$ also grows with $e B$ (see dotted curves in the left panel of the figure). 
When the tadpole term in Eq.(\ref{tr-anom}) is turned on, 
%Fig.~\ref{M-peak}. Since the constituent quark mass 
$M_u$ is significantly lifted up by increasing $e B$ at high temperatures ($T \gtrsim T_{\rm pc}$), 
while gets almost unchanged at low temperatures (see solid curves in the left panel).  
This is due to the additional chiral-explicit breaking effect induced and enhanced by 
the thermomagnetic tadpole term (the second term) in Eq.(\ref{tr-anom}). 
Thus this trend can lead to the reduction of $T_{\rm p c}$ with the increase of $e B$ 
(solid curves in the right panel), 
which is in contrast to the case without the tadpole term in Eq.(\ref{tr-anom}) (dotted curves in the right panel). 
\cred{We have checked that similar trends are observed also for the constituent mass of the down quark, $M_d$, at the physical
point. The constituent mass of the strange quark, $M_s$, behaves with $T$ 
in a way different from $M_u$ and $M_d$, and does not fast damp with $T$, 
simply because of its heaviness, hence is not essential in discussing the (pseudo)criticality of the chiral symmetry restoration. Therefore, it is not the focus of the present paper, to be disregarded in the present discussion. }

\subsection{$T_{\rm p c}$ does not drop with $e B$ in the smaller quark mass range}

\begin{figure}[t] 
\centering 
\includegraphics[width=0.6\textwidth]{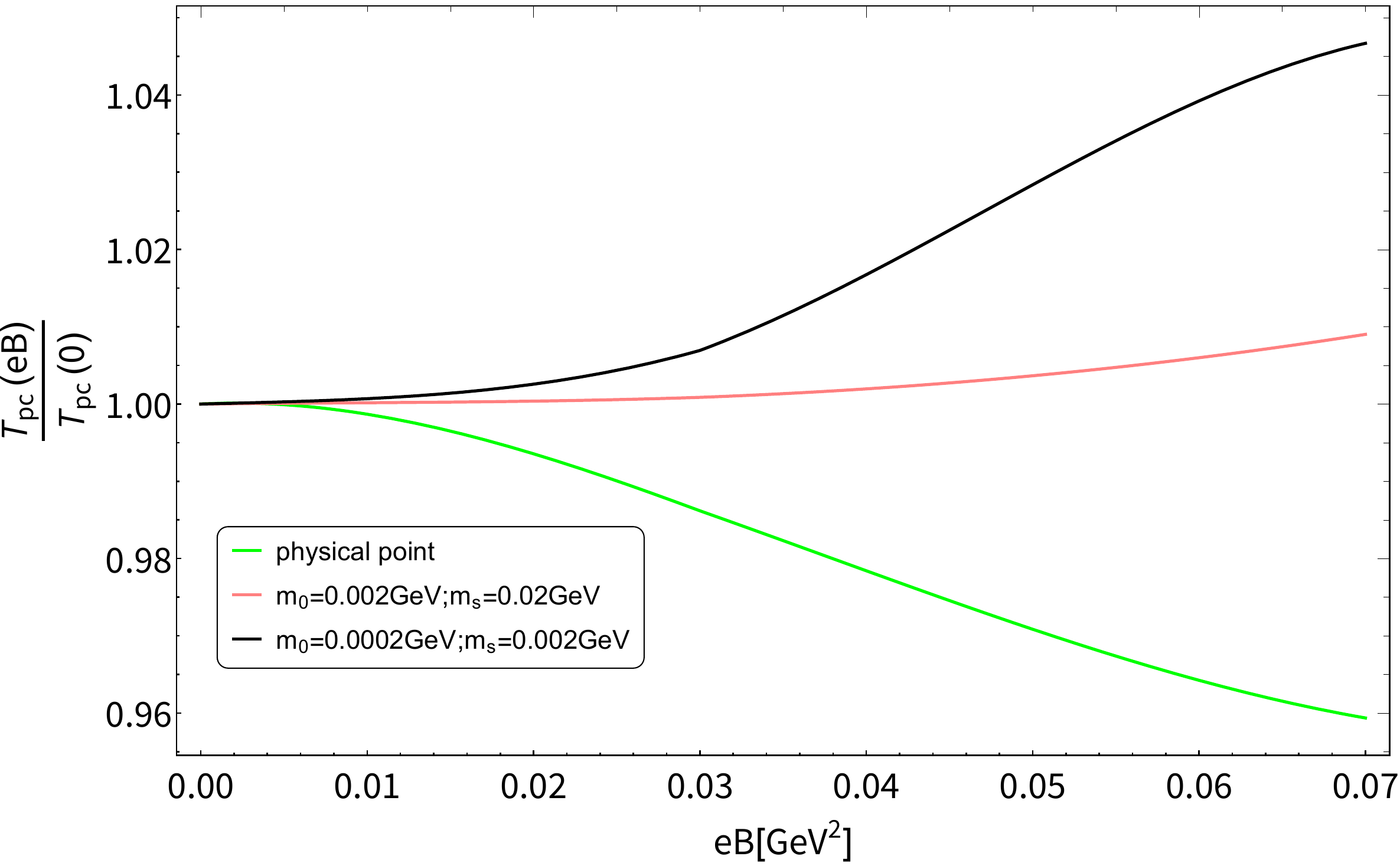}
\caption{The plot on the $eB$ dependence of $T_{\rm p c}$, normalized to $T_{\rm p c}$ at the vacuum ($e B=0$), with varying quark masses. 
%The dashed curve corresponds to the case with $m_0=0.002 \, \mathrm{GeV}$ and 
%$m_s=0.02\, \mathrm{GeV}$. 
%$T_{\rm p c}(e B) / T_{p c}(0) $ keeps almost constant with $e B$.  
$T_{\mathrm{pc}}$ decreases with $eB$ if the quark masses are greater than the critical values, and turns to 
increase when the quark masses gets smaller than the critical ones. 
} 
\label{Tpc}
\end{figure}

%In order to analyze $T_{p c}$ at different quark masses  and thus adjust the quark mass deviation %from the physical point, a total of seven sets of quark mass references are chosen, including the %chiral limit case. 
In Fig.~\ref{Tpc} we show $T_{\rm p c}(eB)$, normalized to $T_{\rm p c}(0)$ ($T_{\rm p c}(eB)$ at $eB = 0$), as a function of $e B$ with various quark masses. 
As clearly seen from the figure, 
$T_{\rm p c}$ does not drop with $e B$ for smaller quark masses 
unlike the case at the physical point, 
rather tends to increase with $e B$ as quark masses get smaller. 
We have observed the critical point around $m_0=0.002\, \mathrm{GeV}$ and $m_s=0.02 \, \mathrm{GeV}$.  
This implies that the expected universal drop trend of $T_{\rm pc}$ 
with the increase $e B$ may not be seen in the small quark mass regime. 
Note that since the thermomagnetic potential in Eq.(\ref{Omega}) is quark-flavor universal up to 
the electromagnetic charge difference, the presently observed trend is almost flavor universal.  
%i.e., the flavor-independent feature seen in the quark mass plane when  in the Columbia plot.   

Trying a smooth and simple extrapolation from lattice results in a strong $eB$ regime to the weak $eB$ regime,  
one could make the present NJL model extended by a running effect associated with 
an intrinsic $eB$ and/or $T$ dependence into the four-fermion coupling $G$ to achieve 
more quantitative agreement between the model and lattice data on the quark condensates 
by the fitting procedure. 
To this end, we may refer to the literature~\cite{Ferreira:2014kpa} and introduce the fitting function of $G$ for the three-flavor model case as   
\begin{equation}
G(B)=G \left( \frac{1 + a\, \xi^2 + b\, \xi^3}{1 + c\, \xi^2 + d\, \xi^4} \right)  
\,,  \label{G-run}
\end{equation}
with $\xi = eB/\Lambda_{\rm QCD}^2$. 
The fitting parameters introduced above are fixed as follows~\cite{Ferreira:2014kpa}:  
\begin{align}
&a = 0.0108805\,, \qquad 
b= - 1.0133 \times 10^{-4} 
\,, \notag\\ 
& c = 0.02228\,, \qquad 
d = 1.84558 \times 10^{-4} 
\,, \label{tab}
\end{align}
(for $\Lambda_{\rm QCD} = 300$ MeV) so that the lattice data on the crossover of the averaged lightest-flavor quark condensate $(\langle \bar{u}u \rangle +\langle \bar{d}d \rangle  )/2$~\cite{Bali:2012zg}, including the inverse magnetic catalysis can even quantitatively be produced well~\cite{Ferreira:2014kpa}. 
It has also been shown that thermodynamic quantities such as 
pressure, entropy density, and equation of state, can show the $eB$ and $T$ dependence 
consistent with the lattice data~\cite{Bali:2014kia}.

\begin{figure}[t] 
\centering 
\includegraphics[width=0.6\textwidth]{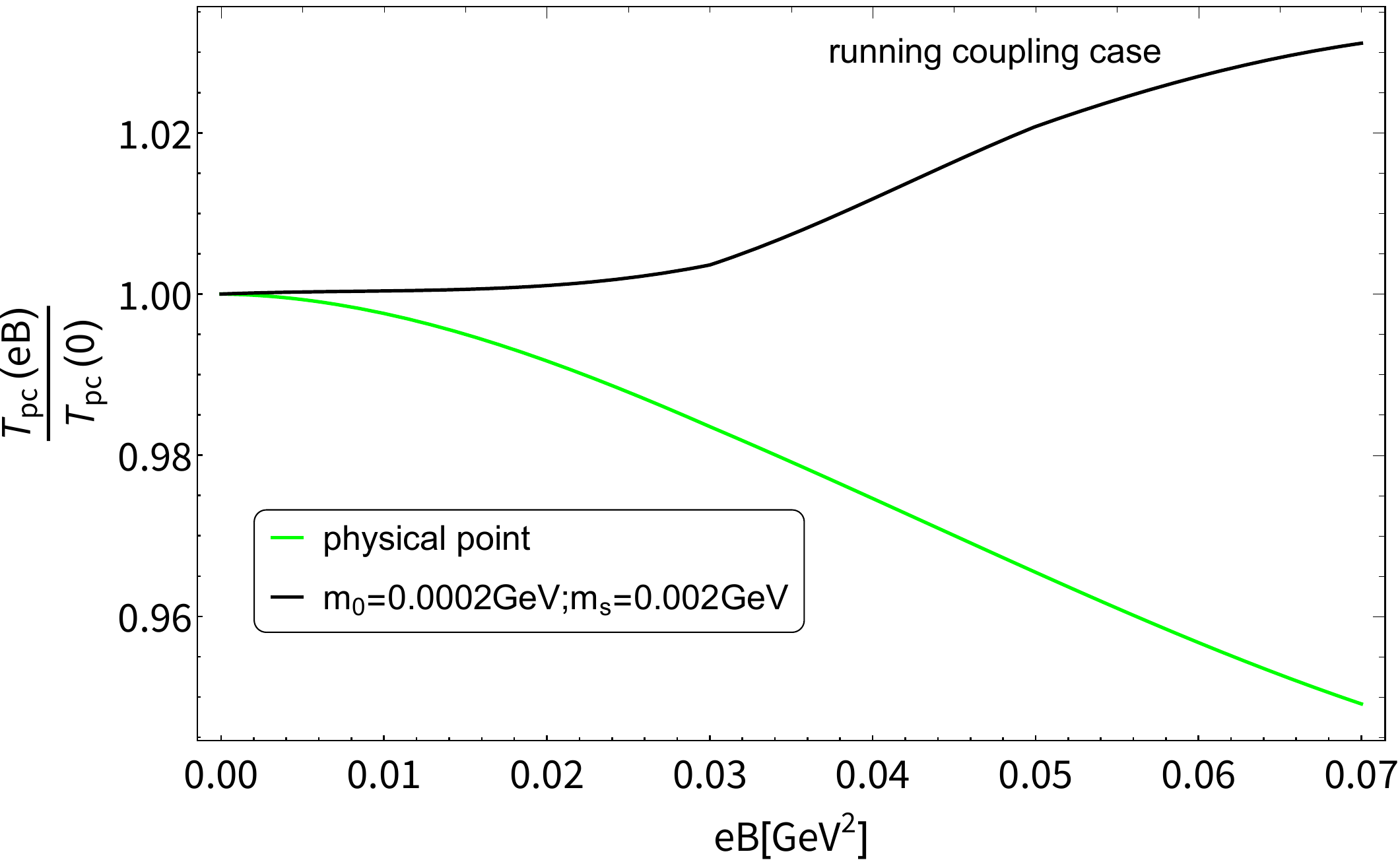}
\caption{
The plots on $T_{\rm pc}$ versus $eB$ varying the quark mass values, 
using   
the running $G$ in Eq.(\ref{G-run}) extrapolated from a strong $eB$ to a weak $eB$ regime based 
on the fitting parameter values listed in Eq.(\ref{tab}). 
The essentially same trend has been observed as predicted in the case without the running effect, in Fig.~\ref{Tpc}, in agreement with the analytic argument in the text: 
the electromagnetic scale anomaly is the crucial contribution to make $T_{\rm pc}$ highly dependent on $m$. 
} 
\label{run-normalized-Tpc}
\end{figure}

When the fitting to the lattice data in the strong $eB$ regime is simply extrapolated to a weaker $eB$ regime ($eB  \lesssim 0.07\,{\rm GeV}^2$ as in Fig.~\ref{Tpc}), 
the running $G$ in Eq.(\ref{G-run}) will almost become constant in $eB$. 
Hence no sizable contribution to trigger the inverse phenomenon will be left in the weak $eB$ regime. Also as to $T_{\rm pc}$, 
the size of the correction is expected to be small enough not to alter 
the qualitative features as has been observed in Fig.~\ref{Tpc}, as long as 
the electromagnetic scale anomaly term is the dominant source to provide the $eB$ dependence with $M$, as in the case with the vacuum place of $M$. 
% (Eq.(\ref{M-VEV})). 
In Fig.~\ref{run-normalized-Tpc} we plot $T_{\rm pc}$ versus $eB$ varying the quark mass values, 
for the cases with  
the running $G$ extrapolated from a strong $eB$ to a weak $eB$ regime based on 
the fitting parameters in Eq.(\ref{tab}). 
The trend of non-dropping $T_{\rm pc}$ still shows up for smaller quark masses in a way qualitatively similar  
to the case with the constant $G$ in Fig.~ \ref{Tpc}. 
At the physical point, the drop rate and grow rates of $T_{\rm pc}$ from $eB=0$ to $eB=0.07\,{\rm GeV}^2$ get smaller by $\sim 1 \%$. 
%while the grow rate in smaller quark mass cases gets smaller by $\sim 2\%$.  

\section{Summary and discussions}

In summary, the drop trend of $T_{\rm pc}$ with respect to the increase of $eB$ may not be 
universal for all the QCD parameter space. 
Employing the 2 + 1 flavor NJL model in the MFA with the electromagnetic scale anomaly included (Eq.(\ref{tr-anom})), 
we have observed that $T_{\rm pc}$  with the weak magnetic field ($eB \lesssim \Lambda_{\rm QCD}^2$)  gets increased in a smaller quark mass regime, which is signaled by the significance of 
the thermomagnetic contribution to the electromagnetic scale anomaly
(See Figs.~\ref{M-peak}, \ref{Tpc}, and \ref{run-normalized-Tpc}).

We have also checked that this non-drop trend will actually be unseen in the case of 
the two-flavor NJL model: $T_{\rm pc}$ decreases with increasing $eB$ even in the small quark mass regime. 
This is simply because of somewhat small contributions to the electromagnetic scale anomaly, compared to the three-flavor case.

Including the Polyakov loop variable into the NJL model,  
the tadpole potential term induced from the electromagnetic scale anomaly would get 
the Polyakov loop dependence via the Fermi-Dirac distribution function part in $F$ in the 
thermomagnetic term (see Eq.(\ref{F})).  
Actually, this extension involves a theoretically nontrivial issue on 
the correlation between the scale anomaly and confinement, hence would provide more phenomenological consequences other than the physics solely on $T_{\rm pc}$.  
This issue is thus to be left to be pursued for another publication.

Lattice simulations have been systematically investigated for 2 + 1 flavors chiral properties in thermal QCD~\cite{HotQCD:2019xnw} and in strong magnetic field~\cite{Tomiya:2019nym,Ding:2020inp,Ding:2020hxw}.  Currently the lower bound of the magnetic field strength that lattice QCD can create at the physical point is $\sim100 \,\mathrm{MeV}$~\cite{Bali:2011qj}. 
The functional renormalization group approach~\cite{Braun:2020ada,Gao:2021vsf} have also developed methods to study the QCD phase diagram in the absence of magnetic field close to the chiral limit. 
We expect that the enhanced trend of $T_{\rm p c}$ with respect to a weak magnetic field in the range of smaller quark masses can be cross-checked by those methods beyond the MFA in the future.

%\item
%In purely thermal QCD, both lattice~\cite{Aoki:2021qws} and chiral effective model~\cite{Meggiolaro:2014eua} results %show that $U(1)_A$ symmetry breaking dominates the contribution of quark condensate as compared to chiral symmetry %breaking. Our result shows that both $T_{p c}$ and quark condensate increase with $e B$ in the smaller quark mass %range.  According to the  chiral Ward identity $\langle\bar{u} u\rangle+\langle\bar{d} d\rangle=-m_0 
%%\chi_\pi$~\cite{GomezNicola:2016ssy,GomezNicola:2017bhm,Kawaguchi:2020qvg}, the susceptibility of pion $\chi_\pi$ %also becomes larger with $e B$, while the order parameter of $U(1)_A$ symmetry breaking can be described by denoting %the difference between the $\chi_\pi$ and $\chi_\delta$ as $\chi_{\pi-\delta} \equiv \chi_\pi-
%\chi_\delta$~\cite{Wang:2021dcy}, where the $\chi_\pi$ is much larger than that of $\chi_\delta$. Therefore, the %magnetic field eventually enhances the $U(1)_A$ symmetry breaking, which may imply that $U(1)_A$ symmetry breaking is %still giving the dominant contribution to quark condensate even in weak magnetic field, and a more detailed %investigation deserves another publication.

The present study reveals that 
depending on the size of the background field $eB$, 
the weakly magnetized QCD plasma with smaller quark masses may not keep the reduction trend on 
the chiral pseudocritical temperature $T_{\rm pc}$ with respect to $eB$. 
\cred{The phase transition becomes first order when $eB \lesssim 0.04\,{\rm GeV}^2$, i.e., $T_{\rm pc} = T_c$, where the electromagnetic scale anomaly 
magnifies $T_c$ by about 2\%, as has been observed in Figs.~\ref{Tpc} and~\ref{run-normalized-Tpc}. 
This enhancement has not been found in the previous study~\cite{YuanyuanWang:2022nds}. 
We shall briefly address phenomenological impacts of this enhancement. }

As has been noted in the discussion around Eq.(\ref{parameter-set}), our present analysis and results  
can be applied also to a wide class of QCD-like theories, such as dark QCD 
and a non-standard cosmological scenario having the supercooled electroweak phase transition~\cite{Witten:1980ez,Iso:2017uuu,Hambye:2018qjv,Sagunski:2023ynd}.  
In the latter case, the electroweak broken vacuum is to be created at temperatures much lower than  
the typical electroweak scale due to the supercooling triggered by some Beyond the Standard Model, 
so that quarks would still keep massless until the dynamical chiral symmetry breaking 
in QCD takes place at around $T = {\cal O}(100\,{\rm MeV})$. 
The vacuum expectation value of the Higgs field is temporarily governed by the dynamical quark condensate -- hence the quarks get smaller current quark masses than the standard ones --, and finally reaches the electroweak scale after the Beyond the Standard Model sector 
finalizes the supercooling (when the associated potential barrier goes away). 
It has recently been argued based on the ladder Schwinger-Dyson approach 
that the QCD chiral phase transition in this type of scenario takes place along with 
keeping all the six quarks light enough at the QCD scale~\cite{Guan:2024ccw}. 
In particular, the QCD-driven Higgs-vacuum expectation value yields $m_t\sim 100$ MeV, and 
others $\lesssim {\cal O}$(MeV) depending on the hierarchical Yukawa couplings, 
and the QCD chiral phase transition is suggested to be first order in the framework of the MFA~\cite{Guan:2024ccw}. 
Given this point, recall that the thermomagnetic part of the electromagnetic scale anomaly term is almost flavor-universal 
and simply scales with the number of light quark flavors, $N_f$ (see Eq.(\ref{tr-anom})). 
Thus, when a redshifted primordial magnetic field is present in the QCD phase transition epoch 
keeping the weak strength enough that the electromagnetic scale anomaly is not washed out, 
the presently gained result implies an increased critical temperature for 
the supercooled electroweak phase transition in the scenarios of this class. 
This might also indicate a larger nucleation/percolation temperature for the supercooling cosmological 
phase transition when the created potential barrier does not substantially get affected by 
the weak $eB$ compared to the case without $eB$. 
This might also affect the gravitational wave production in this type of scenarios. 
\cred{If the nucleation/percolation temperature gets enhanced as well as 
$T_c$ by about 2\% as noted above, the peak frequency of the produced gravitational wave 
would be shifted to higher by the same amount, as long as 
the strength of the magnetic field is small enough not to spoil creation and nucleation of isotropic 
bubbles by the induced inhomogeneity. 
This shift may or may not be tiny or less sensitive to 
the predicted gravitational spectra, which highly depends on modeling 
of the supercooling extension of the Standard Model.  
More on such cosmological implications would be noteworthy to pursue in another publication. Although the quantitative impact is still unclear at this point, at any rate, the present study has proposed a nontrivial indication 
arising from the electromagnetic scale anomaly 
to the prediction related to the gravitational wave production 
or the supercooling phase transition. 
}

Regarding the quark mass and $eB$ dependence in light of 
the impact on the cosmological dark QCD scenarios, note that 
%The current analysis has been done based on the NJL model 
%in the mean field approximation. 
%the presently applied approach is on the same level of reliability as in the literature
the method currently employed is comparable in reliability to those discussed in references~\cite{Holthausen:2013ota,Ametani:2015jla,Aoki:2017aws,Helmboldt:2019pan,Reichert:2021cvs,Aoki:2019mlt}.
%in a sense of employing low-energy effective models to address the chiral 
%phase transition instead of the underlying full dark QCD. 
%These methods investigate the chiral phase transition by using low-energy effective models rather than the %underlying full dark QCD.
%Since the linear sigma model belongs to the same universality class 
%as NJL-like models in terms of the renormalization group, 
%the presently deduced conclusion 
%could directly be applied also to the other dark QCD scenario
%~\cite{Heikinheimo:2018esa,Bai:2018dxf,Archer-Smith:2019gzq,Dvali:2019ewm,Easa:2022vcw,Helmboldt:2019pan,Reichert:2021cvs,Tsumura:2017knk,Croon:2019iuh} argued based on the linear-sigma model description including 
%the pioneer work by Pisarski and Wilczek~\cite{Pisarski:1983ms}. 
As the Wilsonian renormalization group tells that 
the linear sigma model belongs to the same universality class as NJL-like models, 
the presently derived conclusion could also be directly extended to other dark QCD scenarios~\cite{Heikinheimo:2018esa,Bai:2018dxf,Archer-Smith:2019gzq,Dvali:2019ewm,Easa:2022vcw,Helmboldt:2019pan,Reichert:2021cvs,Tsumura:2017knk,Croon:2019iuh} based on the linear sigma model analysis like done in the original Pisarski and Wilczek~\cite{Pisarski:1983ms}.
In this sense, we may say that the present study has an adequate impact as the first step. 
The analysis should anyhow be improved by going beyond the MFA, which is made possible, say, by 
using the functional renormalization group method.  
%Although the first step is sufficiently effective, the current model analysis should in any case be %enhanced by going beyond MFA (e.g., functional renormalization group method) to make it more rigorous.
It would help get a more conclusive answer to whether 
%the first-order nature of 
%the chiral phase transition surely goes away in 
%thermomagnetic QCD with massless three flavors.  
%In this way, we would get a more conclusive answer to the question of whether 
the chiral $T_{pc}$ in thermomagnetic QCD drops with $eB$ in the smaller quark mass regime.

\section*{Acknowledgments} 

We thank Maxim Chernodub for useful comments. 
This work was supported in part by the National Science Foundation of China (NSFC) under Grant No.11747308, 11975108, 12047569, 
and the Seeds Funding of Jilin University (S.M.).  
The work by M.K. was supported by the Fundamental Research Funds for the Central Universities 
and partially by the National Natural Science Foundation of China (NSFC) Grant  No. 12235016, and the Strategic Priority Research Program of Chinese Academy of Sciences under Grant No. XDB34030000.
The work of A.T. was partially supported by JSPS  KAKENHI Grant Numbers 20K14479, 22K03539, 22H05112, and 22H05111, and MEXT as ``Program for Promoting Researches on the Supercomputer Fugaku'' (Simulation for basic science: approaching the new quantum era; Grant Number JPMXP1020230411, and 
Search for physics beyond the standard model using large-scale lattice QCD simulation and development of AI technology toward next-generation lattice QCD; Grant Number JPMXP1020230409).

\appendix

\section{Computation of the electromagnetic scale anomaly at fermion-one-loop level} 
\label{app}

As was discussed in the literature~\cite{Kawaguchi:2020kce}, 
we apply the so-called ``Higgs low-energy theorem"~\cite{Shifman:1979eb} to the chiral-singlet 
component of the quarkonic mesons in QCD, $\varphi$, 
so that the trace anomaly coupled to $\varphi$  
is evaluated in the {\it soft-dilaton limit} as~\footnote{See also the literature~\cite{Matsuzaki:2012vc}, 
for the discussion on 
the Ward-Takahashi identities for the scale symmetry breaking to get a composite 
dilaton coupling to gauge bosons.}
\begin{eqnarray}
{\cal L }_{\varphi-A^{\rm em}-A^{\rm em}} 
= 
%\frac{\varphi}{f_\varphi} \left[ 
%\frac{\beta (e)}{2e} F_{\mu\nu}^2 \right] 
%\nonumber\\
%&=&
\frac{\varphi}{f_\varphi} \left[
\left( 
\lim_{q \to0} q^\lambda \frac{\partial }{\partial q^\lambda} \Pi(q)
\right) 
\frac{1}{4} F_{\mu\nu}^2 
\right] 
%\notag\\ 
%%&=&
%\frac{\varphi}{f_\varphi} \left[
%\left( 
%\lim_{q\to0} m\frac{\partial }{\partial m} \Pi(q)
%\right) 
%\frac{1}{4} F_{\mu\nu}^2 
%\right] 
\,. \label{soft-dilaton}
\end{eqnarray}
Here $F_{\mu\nu}$ is the field strength of the photon field ($A_\mu^{\rm em}$), and 
$\Pi(q)$ denotes the photon polarization correlator associated with the  
the photon polarization tensor $\Pi_{\mu\nu}(q)$,  
\begin{eqnarray}
\Pi^{\mu\nu}(q) &=&
i (q^2 g^{\mu\nu}- q^\mu q^\nu) \Pi(q) 
\,,  
\end{eqnarray}
with the transfer momentum $q$ (which is $\sim m_\varphi \sim 0$ in the soft-dilaton approximation). 
We evaluate $\lim_{q\to0} q^\lambda \frac{\partial }{\partial q^\lambda} \Pi(q)$ 
at the quark-one loop level. 
Since $\varphi$ acts as a quark mass $m$ in the quark propagator and $\lim_{q\to 0}q^\mu\frac{\partial }{\partial q^\mu }\Pi(q)$ is nothing but the first order term in the Taylor expansion coefficient with respect to $\varphi$ for the photon-wavefunction renormalization function, i.e., $\Pi(q^2)$, we have 
\begin{eqnarray}
\lim_{q\to 0}q^\mu\frac{\partial }{\partial q^\mu }\Pi(q)
=
\lim_{q\to0} m\frac{\partial }{\partial m} \Pi(q)
\,.   
\end{eqnarray}
Working in the imaginary time formalism, we thus compute 
$\lim_{q\to0} m\frac{\partial }{\partial m} \Pi(q)$ to find 
\begin{eqnarray}
\lim_{q\to0} m\frac{\partial }{\partial m} \Pi(q)&=&
%16e^2 N_c 
%m^2
%\int_0^1 dx
%\left\{
%\left(
%x -x^2
%\right)
%\right\}
%2i\int \frac{d^4l}{(2\pi)^4}\frac{1}{(l^2 -m^2)^3}\nonumber\\
%&=&
%16e^2 N_c 
%\frac{i m^2}{3}
%\int \frac{d^4l}{(2\pi)^4}\frac{1}{(l^2 -m^2)^3}\nonumber\\
%&\to&
%16e^2 N_c 
%\frac{i m^2}{3}
%iT\sum_n \int \frac{d^3l}{(2\pi)^3}\frac{1}{(-\omega_n^2 -m^2-\vec l^2)^3}\nonumber\\
%&=&
16e^2 N_c 
\frac{ m^2}{3}
T\sum_N \int \frac{d^3l}{(2\pi)^3}\frac{1}{(\omega_N^2 +E_l^2)^3}\nonumber\\
&=&
16e^2 N_c 
\frac{ m^2}{3}
 \int \frac{d^3l}{(2\pi)^3}
 \frac{3}{16E_l^5}(1-2f(E_l))
\nonumber\\
%&=&
%e^2m^2 N_c  
% \int \frac{d^3l}{(2\pi)^3}
% \frac{1}{E_l^5}(1-2f(E_l))\nonumber\\
 &=&
 \frac{6 e^2 N_c }{4(4 \pi)^2} +
e^2m^2  N_c  
 \int \frac{d^3l}{(2\pi)^3}
 \frac{1}{E_l^5}(-2f(E_l))\nonumber\\ 
  &=&
 \frac{2\beta(e) }{e} +
e^2m^2  N_c    
 \int \frac{d^3l}{(2\pi)^3}
 \frac{1}{E_l^5}(-2f(E_l)) 
 \,, 
\end{eqnarray}
where $f(E_l)$ is the Fermi-Dirac distribution function, $f(E_l)= 1/(1 + e^{E_l/T})$ with 
$E_l = \sqrt{\vec{l}^2 + m^2}$. 
For simplicity, we have taken the electromagnetic charge of the fermion $q_f=1$. 
Passing through the replacement of the three-momentum integration with 
the sum of the Landau levels and the residual one-dimensional integration in the $z$-direction, 
one can arrive at the expression in the right hand side of Eq.(\ref{tr-anom}) for each 
quark contribution with the charge of $q_f$.


\begin{thebibliography}{99} 


%\cite{Bali:2011qj}
\bibitem{Bali:2011qj}
G.~S.~Bali, F.~Bruckmann, G.~Endrodi, Z.~Fodor, S.~D.~Katz, S.~Krieg, A.~Schafer and K.~K.~Szabo,
%``The QCD phase diagram for external magnetic fields,''
JHEP \textbf{02}, 044 (2012)
doi:10.1007/JHEP02(2012)044
[arXiv:1111.4956 [hep-lat]].
%754 citations counted in INSPIRE as of 13 May 2025

%\cite{Bornyakov:2013eya}
\bibitem{Bornyakov:2013eya}
V.~G.~Bornyakov, P.~V.~Buividovich, N.~Cundy, O.~A.~Kochetkov and A.~Sch\"afer,
%``Deconfinement transition in two-flavor lattice QCD with dynamical overlap fermions in an external magnetic field,''
Phys. Rev. D \textbf{90}, no.3, 034501 (2014)
doi:10.1103/PhysRevD.90.034501
[arXiv:1312.5628 [hep-lat]].
%113 citations counted in INSPIRE as of 29 Apr 2025

%\cite{Bali:2014kia}
\bibitem{Bali:2014kia}
G.~S.~Bali, F.~Bruckmann, G.~Endr\"odi, S.~D.~Katz and A.~Sch\"afer,
%``The QCD equation of state in background magnetic fields,''
JHEP \textbf{08}, 177 (2014)
doi:10.1007/JHEP08(2014)177
[arXiv:1406.0269 [hep-lat]].
%269 citations counted in INSPIRE as of 03 May 2025

%\cite{Tomiya:2019nym}
\bibitem{Tomiya:2019nym}
A.~Tomiya, H.~T.~Ding, X.~D.~Wang, Y.~Zhang, S.~Mukherjee and C.~Schmidt,
%``Phase structure of three flavor QCD in external magnetic fields using HISQ fermions,''
PoS \textbf{LATTICE2018}, 163 (2019)
doi:10.22323/1.334.0163
[arXiv:1904.01276 [hep-lat]].
%22 citations counted in INSPIRE as of 17 Apr 2025

%\cite{DElia:2018xwo}
\bibitem{DElia:2018xwo}
M.~D'Elia, F.~Manigrasso, F.~Negro and F.~Sanfilippo,
%``QCD phase diagram in a magnetic background for different values of the pion mass,''
Phys. Rev. D \textbf{98}, no.5, 054509 (2018)
doi:10.1103/PhysRevD.98.054509
[arXiv:1808.07008 [hep-lat]].
%93 citations counted in INSPIRE as of 22 Apr 2025

%\cite{Endrodi:2019zrl}
\bibitem{Endrodi:2019zrl}
G.~Endrodi, M.~Giordano, S.~D.~Katz, T.~G.~Kov\'acs and F.~Pittler,
%``Magnetic catalysis and inverse catalysis for heavy pions,''
JHEP \textbf{07}, 007 (2019)
doi:10.1007/JHEP07(2019)007
[arXiv:1904.10296 [hep-lat]].
%63 citations counted in INSPIRE as of 17 Apr 2025

%\cite{Vachaspati:1991nm}
\bibitem{Vachaspati:1991nm}
T.~Vachaspati,
%``Magnetic fields from cosmological phase transitions,''
Phys. Lett. B \textbf{265}, 258-261 (1991)
doi:10.1016/0370-2693(91)90051-Q
%763 citations counted in INSPIRE as of 28 Apr 2025

%\cite{Enqvist:1993np}
\bibitem{Enqvist:1993np}
K.~Enqvist and P.~Olesen,
%``On primordial magnetic fields of electroweak origin,''
Phys. Lett. B \textbf{319}, 178-185 (1993)
doi:10.1016/0370-2693(93)90799-N
[arXiv:hep-ph/9308270 [hep-ph]].
%184 citations counted in INSPIRE as of 28 Apr 2025

%\cite{Grasso:1997nx}
\bibitem{Grasso:1997nx}
D.~Grasso and A.~Riotto,
%``On the nature of the magnetic fields generated during the electroweak phase transition,''
Phys. Lett. B \textbf{418}, 258-265 (1998)
doi:10.1016/S0370-2693(97)01224-0
[arXiv:hep-ph/9707265 [hep-ph]].
%75 citations counted in INSPIRE as of 05 May 2025

%\cite{Grasso:2000wj}
\bibitem{Grasso:2000wj}
D.~Grasso and H.~R.~Rubinstein,
%``Magnetic fields in the early universe,''
Phys. Rept. \textbf{348}, 163-266 (2001)
doi:10.1016/S0370-1573(00)00110-1
[arXiv:astro-ph/0009061 [astro-ph]].
%1034 citations counted in INSPIRE as of 05 May 2025

%\cite{Ellis:2019tjf}
\bibitem{Ellis:2019tjf}
J.~Ellis, M.~Fairbairn, M.~Lewicki, V.~Vaskonen and A.~Wickens,
%``Intergalactic Magnetic Fields from First-Order Phase Transitions,''
JCAP \textbf{09}, 019 (2019)
doi:10.1088/1475-7516/2019/09/019
[arXiv:1907.04315 [astro-ph.CO]].
%56 citations counted in INSPIRE as of 29 Apr 2025

%\cite{Zhang:2019vsb}
\bibitem{Zhang:2019vsb}
Y.~Zhang, T.~Vachaspati and F.~Ferrer,
%``Magnetic field production at a first-order electroweak phase transition,''
Phys. Rev. D \textbf{100}, no.8, 083006 (2019)
doi:10.1103/PhysRevD.100.083006
[arXiv:1902.02751 [hep-ph]].
%30 citations counted in INSPIRE as of 17 Apr 2025

%\cite{Di:2020kbw}
\bibitem{Di:2020kbw}
Y.~Di, J.~Wang, R.~Zhou, L.~Bian, R.~G.~Cai and J.~Liu,
%``Magnetic Field and Gravitational Waves from the First-Order Phase Transition,''
Phys. Rev. Lett. \textbf{126}, no.25, 251102 (2021)
doi:10.1103/PhysRevLett.126.251102
[arXiv:2012.15625 [astro-ph.CO]].
%51 citations counted in INSPIRE as of 06 May 2025

%\cite{Yang:2021uid}
\bibitem{Yang:2021uid}
J.~Yang and L.~Bian,
%``Magnetic field generation from bubble collisions during first-order phase transition,''
Phys. Rev. D \textbf{106}, no.2, 023510 (2022)
doi:10.1103/PhysRevD.106.023510
[arXiv:2102.01398 [astro-ph.CO]].
%20 citations counted in INSPIRE as of 17 Apr 2025

%\cite{Turner:1987vd}
\bibitem{Turner:1987vd}
M.~S.~Turner and L.~M.~Widrow,
%``Gravitational Production of Scalar Particles in Inflationary Universe Models,''
Phys. Rev. D \textbf{37}, 3428 (1988)
doi:10.1103/PhysRevD.37.3428
%42 citations counted in INSPIRE as of 09 May 2025

%\cite{Garretson:1992vt}
\bibitem{Garretson:1992vt}
W.~D.~Garretson, G.~B.~Field and S.~M.~Carroll,
%``Primordial magnetic fields from pseudoGoldstone bosons,''
Phys. Rev. D \textbf{46}, 5346-5351 (1992)
doi:10.1103/PhysRevD.46.5346
[arXiv:hep-ph/9209238 [hep-ph]].
%334 citations counted in INSPIRE as of 18 Apr 2025

%\cite{Anber:2006xt}
\bibitem{Anber:2006xt}
M.~M.~Anber and L.~Sorbo,
%``N-flationary magnetic fields,''
JCAP \textbf{10}, 018 (2006)
doi:10.1088/1475-7516/2006/10/018
[arXiv:astro-ph/0606534 [astro-ph]].
%252 citations counted in INSPIRE as of 29 Apr 2025

%\cite{Domcke:2019mnd}
\bibitem{Domcke:2019mnd}
V.~Domcke, B.~von Harling, E.~Morgante and K.~Mukaida,
%``Baryogenesis from axion inflation,''
JCAP \textbf{10}, 032 (2019)
doi:10.1088/1475-7516/2019/10/032
[arXiv:1905.13318 [hep-ph]].
%87 citations counted in INSPIRE as of 17 Apr 2025

%\cite{Domcke:2019qmm}
\bibitem{Domcke:2019qmm}
V.~Domcke, Y.~Ema and K.~Mukaida,
%``Chiral Anomaly, Schwinger Effect, Euler-Heisenberg Lagrangian, and application to axion inflation,''
JHEP \textbf{02}, 055 (2020)
doi:10.1007/JHEP02(2020)055
[arXiv:1910.01205 [hep-ph]].
%72 citations counted in INSPIRE as of 17 Apr 2025

%\cite{Patel:2019isj}
\bibitem{Patel:2019isj}
T.~Patel, H.~Tashiro and Y.~Urakawa,
%``Resonant magnetogenesis from axions,''
JCAP \textbf{01}, 043 (2020)
doi:10.1088/1475-7516/2020/01/043
[arXiv:1909.00288 [astro-ph.CO]].
%30 citations counted in INSPIRE as of 10 May 2025

%\cite{Domcke:2020zez}
\bibitem{Domcke:2020zez}
V.~Domcke, V.~Guidetti, Y.~Welling and A.~Westphal,
%``Resonant backreaction in axion inflation,''
JCAP \textbf{09}, 009 (2020)
doi:10.1088/1475-7516/2020/09/009
[arXiv:2002.02952 [astro-ph.CO]].
%81 citations counted in INSPIRE as of 01 May 2025

%\cite{Shtanov:2020gjp}
\bibitem{Shtanov:2020gjp}
Y.~Shtanov and M.~Pavliuk,
%``Model-independent constraints in inflationary magnetogenesis,''
JCAP \textbf{08}, 042 (2020)
doi:10.1088/1475-7516/2020/08/042
[arXiv:2004.00947 [astro-ph.CO]].
%18 citations counted in INSPIRE as of 17 Apr 2025

%\cite{Okano:2020uyr}
\bibitem{Okano:2020uyr}
S.~Okano and T.~Fujita,
%``Chiral Gravitational Waves Produced in a Helical Magnetogenesis Model,''
JCAP \textbf{03}, 026 (2021)
doi:10.1088/1475-7516/2021/03/026
[arXiv:2005.13833 [astro-ph.CO]].
%24 citations counted in INSPIRE as of 23 Apr 2025

%\cite{Cado:2021bia}
\bibitem{Cado:2021bia}
Y.~Cado, B.~von Harling, E.~Mass\'o and M.~Quir\'os,
%``Baryogenesis via gauge field production from a relaxing Higgs,''
JCAP \textbf{07}, 049 (2021)
doi:10.1088/1475-7516/2021/07/049
[arXiv:2102.13650 [hep-ph]].
%11 citations counted in INSPIRE as of 10 May 2025

%\cite{Kushwaha:2021csq}
\bibitem{Kushwaha:2021csq}
A.~Kushwaha and S.~Shankaranarayanan,
%``Helical magnetic fields from Riemann coupling lead to baryogenesis,''
Phys. Rev. D \textbf{104}, no.6, 063502 (2021)
doi:10.1103/PhysRevD.104.063502
[arXiv:2103.05339 [hep-ph]].
%14 citations counted in INSPIRE as of 10 May 2025

%\cite{Gorbar:2021rlt}
\bibitem{Gorbar:2021rlt}
E.~V.~Gorbar, K.~Schmitz, O.~O.~Sobol and S.~I.~Vilchinskii,
%``Gauge-field production during axion inflation in the gradient expansion formalism,''
Phys. Rev. D \textbf{104}, no.12, 123504 (2021)
doi:10.1103/PhysRevD.104.123504
[arXiv:2109.01651 [hep-ph]].
%67 citations counted in INSPIRE as of 10 May 2025

%\cite{Gorbar:2021zlr}
\bibitem{Gorbar:2021zlr}
E.~V.~Gorbar, K.~Schmitz, O.~O.~Sobol and S.~I.~Vilchinskii,
%``Hypermagnetogenesis from axion inflation: Model-independent estimates,''
Phys. Rev. D \textbf{105}, no.4, 043530 (2022)
doi:10.1103/PhysRevD.105.043530
[arXiv:2111.04712 [hep-ph]].
%30 citations counted in INSPIRE as of 17 Apr 2025

%\cite{Gorbar:2021ajq}
\bibitem{Gorbar:2021ajq}
E.~V.~Gorbar, A.~I.~Momot, I.~V.~Rudenok, O.~O.~Sobol, S.~I.~Vilchinskii and I.~V.~Oleinikova,
%``Chirality Production during Axion Inflation,''
Ukr. J. Phys. \textbf{68}, no.11, 717 (2023)
doi:10.15407/ujpe68.11.717
[arXiv:2111.05848 [hep-ph]].
%7 citations counted in INSPIRE as of 17 Apr 2025

%\cite{Fujita:2022fwc}
\bibitem{Fujita:2022fwc}
T.~Fujita, J.~Kume, K.~Mukaida and Y.~Tada,
%``Effective treatment of U(1) gauge field and charged particles in axion inflation,''
JCAP \textbf{09}, 023 (2022)
doi:10.1088/1475-7516/2022/09/023
[arXiv:2204.01180 [hep-ph]].
%26 citations counted in INSPIRE as of 17 Apr 2025

%\cite{Bandyopadhyay:2020zte}
\bibitem{Bandyopadhyay:2020zte}
A.~Bandyopadhyay and R.~L.~S.~Farias,
%``Inverse magnetic catalysis: how much do we know about?,''
Eur. Phys. J. ST \textbf{230}, no.3, 719-728 (2021)
doi:10.1140/epjs/s11734-021-00023-1
[arXiv:2003.11054 [hep-ph]].
%45 citations counted in INSPIRE as of 17 Apr 2025

%\cite{Kawaguchi:2021nsa}
\bibitem{Kawaguchi:2021nsa}
M.~Kawaguchi, S.~Matsuzaki and A.~Tomiya,
%``Detecting scale anomaly in chiral phase transition of QCD: new critical endpoint pinned down,''
JHEP \textbf{12}, 175 (2021)
doi:10.1007/JHEP12(2021)175
[arXiv:2102.05294 [hep-ph]].
%7 citations counted in INSPIRE as of 17 Apr 2025

%\cite{YuanyuanWang:2022nds}
\bibitem{YuanyuanWang:2022nds}
Y.~Wang, M.~Kawaguchi, S.~Matsuzaki and A.~Tomiya,
%``Fate of the first-order chiral phase transition in QCD: Implications for dark QCD studied via a Nambu\textendash{}Jona-Lasinio model,''
Phys. Rev. D \textbf{106}, no.9, 095010 (2022)
doi:10.1103/PhysRevD.106.095010
[arXiv:2208.03975 [hep-ph]].
%3 citations counted in INSPIRE as of 17 Apr 2025

%\cite{Dini:2021hug}
\bibitem{Dini:2021hug}
L.~Dini, P.~Hegde, F.~Karsch, A.~Lahiri, C.~Schmidt and S.~Sharma,
%``Chiral phase transition in three-flavor QCD from lattice QCD,''
Phys. Rev. D \textbf{105}, no.3, 034510 (2022)
doi:10.1103/PhysRevD.105.034510
[arXiv:2111.12599 [hep-lat]].
%51 citations counted in INSPIRE as of 17 Apr 2025

%\cite{Bazavov:2017xul}
\bibitem{Bazavov:2017xul}
A.~Bazavov, H.~T.~Ding, P.~Hegde, F.~Karsch, E.~Laermann, S.~Mukherjee, P.~Petreczky and C.~Schmidt,
%``Chiral phase structure of three flavor QCD at vanishing baryon number density,''
Phys. Rev. D \textbf{95}, no.7, 074505 (2017)
doi:10.1103/PhysRevD.95.074505
[arXiv:1701.03548 [hep-lat]].
%77 citations counted in INSPIRE as of 17 Apr 2025

%\cite{Cuteri:2021ikv}
\bibitem{Cuteri:2021ikv}
F.~Cuteri, O.~Philipsen and A.~Sciarra,
%``On the order of the QCD chiral phase transition for different numbers of quark flavours,''
JHEP \textbf{11}, 141 (2021)
doi:10.1007/JHEP11(2021)141
[arXiv:2107.12739 [hep-lat]].
%83 citations counted in INSPIRE as of 17 Apr 2025

%\cite{Holthausen:2013ota}
\bibitem{Holthausen:2013ota}
M.~Holthausen, J.~Kubo, K.~S.~Lim and M.~Lindner,
%``Electroweak and Conformal Symmetry Breaking by a Strongly Coupled Hidden Sector,''
JHEP \textbf{12}, 076 (2013)
doi:10.1007/JHEP12(2013)076
[arXiv:1310.4423 [hep-ph]].
%142 citations counted in INSPIRE as of 17 Apr 2025

%\cite{Ametani:2015jla}
\bibitem{Ametani:2015jla}
Y.~Ametani, M.~Aoki, H.~Goto and J.~Kubo,
%``Nambu-Goldstone Dark Matter in a Scale Invariant Bright Hidden Sector,''
Phys. Rev. D \textbf{91}, no.11, 115007 (2015)
doi:10.1103/PhysRevD.91.115007
[arXiv:1505.00128 [hep-ph]].
%44 citations counted in INSPIRE as of 17 Apr 2025

%\cite{Aoki:2017aws}
\bibitem{Aoki:2017aws}
M.~Aoki, H.~Goto and J.~Kubo,
%``Gravitational Waves from Hidden QCD Phase Transition,''
Phys. Rev. D \textbf{96}, no.7, 075045 (2017)
doi:10.1103/PhysRevD.96.075045
[arXiv:1709.07572 [hep-ph]].
%61 citations counted in INSPIRE as of 24 Apr 2025

%\cite{Bai:2018dxf}
\bibitem{Bai:2018dxf}
Y.~Bai, A.~J.~Long and S.~Lu,
%``Dark Quark Nuggets,''
Phys. Rev. D \textbf{99}, no.5, 055047 (2019)
doi:10.1103/PhysRevD.99.055047
[arXiv:1810.04360 [hep-ph]].
%173 citations counted in INSPIRE as of 10 May 2025

%\cite{Helmboldt:2019pan}
\bibitem{Helmboldt:2019pan}
A.~J.~Helmboldt, J.~Kubo and S.~van der Woude,
%``Observational prospects for gravitational waves from hidden or dark chiral phase transitions,''
Phys. Rev. D \textbf{100}, no.5, 055025 (2019)
doi:10.1103/PhysRevD.100.055025
[arXiv:1904.07891 [hep-ph]].
%92 citations counted in INSPIRE as of 30 Apr 2025

%\cite{Reichert:2021cvs}
\bibitem{Reichert:2021cvs}
M.~Reichert, F.~Sannino, Z.~W.~Wang and C.~Zhang,
%``Dark confinement and chiral phase transitions: gravitational waves vs matter representations,''
JHEP \textbf{01}, 003 (2022)
doi:10.1007/JHEP01(2022)003
[arXiv:2109.11552 [hep-ph]].
%55 citations counted in INSPIRE as of 12 May 2025

%\cite{Archer-Smith:2019gzq}
\bibitem{Archer-Smith:2019gzq}
P.~Archer-Smith, D.~Linthorne and D.~Stolarski,
%``Gravitational Wave Signals from Multiple Hidden Sectors,''
Phys. Rev. D \textbf{101}, no.9, 095016 (2020)
doi:10.1103/PhysRevD.101.095016
[arXiv:1910.02083 [hep-ph]].
%24 citations counted in INSPIRE as of 08 May 2025

%\cite{Aoki:2019mlt}
\bibitem{Aoki:2019mlt}
M.~Aoki and J.~Kubo,
%``Gravitational waves from chiral phase transition in a conformally extended standard model,''
JCAP \textbf{04}, 001 (2020)
doi:10.1088/1475-7516/2020/04/001
[arXiv:1910.05025 [hep-ph]].
%37 citations counted in INSPIRE as of 30 Apr 2025

%\cite{Dvali:2019ewm}
\bibitem{Dvali:2019ewm}
G.~Dvali, E.~Koutsangelas and F.~Kuhnel,
%``Compact Dark Matter Objects via $N$ Dark Sectors,''
Phys. Rev. D \textbf{101}, 083533 (2020)
doi:10.1103/PhysRevD.101.083533
[arXiv:1911.13281 [astro-ph.CO]].
%22 citations counted in INSPIRE as of 05 May 2025

%\cite{Easa:2022vcw}
\bibitem{Easa:2022vcw}
H.~Easa, T.~Gregoire, D.~Stolarski and C.~Cosme,
%``Baryogenesis and dark matter in multiple hidden sectors,''
Phys. Rev. D \textbf{109}, no.7, 075003 (2024)
doi:10.1103/PhysRevD.109.075003
[arXiv:2206.11314 [hep-ph]].
%8 citations counted in INSPIRE as of 10 May 2025

%\cite{Tsumura:2017knk}
\bibitem{Tsumura:2017knk}
K.~Tsumura, M.~Yamada and Y.~Yamaguchi,
%``Gravitational wave from dark sector with dark pion,''
JCAP \textbf{07}, 044 (2017)
doi:10.1088/1475-7516/2017/07/044
[arXiv:1704.00219 [hep-ph]].
%61 citations counted in INSPIRE as of 24 Apr 2025

%\cite{Witten:1980ez}
\bibitem{Witten:1980ez}
E.~Witten,
%``Cosmological Consequences of a Light Higgs Boson,''
Nucl. Phys. B \textbf{177}, 477-488 (1981)
doi:10.1016/0550-3213(81)90182-6
%243 citations counted in INSPIRE as of 17 Apr 2025

%\cite{Iso:2017uuu}
\bibitem{Iso:2017uuu}
S.~Iso, P.~D.~Serpico and K.~Shimada,
%``QCD-Electroweak First-Order Phase Transition in a Supercooled Universe,''
Phys. Rev. Lett. \textbf{119}, no.14, 141301 (2017)
doi:10.1103/PhysRevLett.119.141301
[arXiv:1704.04955 [hep-ph]].
%137 citations counted in INSPIRE as of 17 Apr 2025

%\cite{Hambye:2018qjv}
\bibitem{Hambye:2018qjv}
T.~Hambye, A.~Strumia and D.~Teresi,
%``Super-cool Dark Matter,''
JHEP \textbf{08}, 188 (2018)
doi:10.1007/JHEP08(2018)188
[arXiv:1805.01473 [hep-ph]].
%121 citations counted in INSPIRE as of 21 Apr 2025

%\cite{Sagunski:2023ynd}
\bibitem{Sagunski:2023ynd}
L.~Sagunski, P.~Schicho and D.~Schmitt,
%``Supercool exit: Gravitational waves from QCD-triggered conformal symmetry breaking,''
Phys. Rev. D \textbf{107}, no.12, 123512 (2023)
doi:10.1103/PhysRevD.107.123512
[arXiv:2303.02450 [hep-ph]].
%34 citations counted in INSPIRE as of 12 May 2025

%\cite{Ali:2024mnn}
\bibitem{Ali:2024mnn}
M.~S.~Ali, C.~A.~Islam and R.~Sharma,
%``QCD phase diagram in the T\textendash{}eB plane for varying pion mass,''
Phys. Rev. D \textbf{110}, no.9, 096011 (2024)
doi:10.1103/PhysRevD.110.096011
[arXiv:2407.14449 [hep-ph]].
%6 citations counted in INSPIRE as of 17 Apr 2025

%\cite{Brown:1990ev}
\bibitem{Brown:1990ev}
F.~R.~Brown, F.~P.~Butler, H.~Chen, N.~H.~Christ, Z.~h.~Dong, W.~Schaffer, L.~I.~Unger and A.~Vaccarino,
%``On the existence of a phase transition for QCD with three light quarks,''
Phys. Rev. Lett. \textbf{65}, 2491-2494 (1990)
doi:10.1103/PhysRevLett.65.2491
%526 citations counted in INSPIRE as of 07 May 2025

%\cite{Kobayashi:1970ji}
\bibitem{Kobayashi:1970ji}
M.~Kobayashi and T.~Maskawa,
%``Chiral symmetry and eta-x mixing,''
Prog. Theor. Phys. \textbf{44}, 1422-1424 (1970)
doi:10.1143/PTP.44.1422
%189 citations counted in INSPIRE as of 29 Apr 2025

%\cite{Kobayashi:1971qz}
\bibitem{Kobayashi:1971qz}
M.~Kobayashi, H.~Kondo and T.~Maskawa,
%``Symmetry breaking of the chiral u(3) x u(3) and the quark model,''
Prog. Theor. Phys. \textbf{45}, 1955-1959 (1971)
doi:10.1143/PTP.45.1955
%137 citations counted in INSPIRE as of 29 Apr 2025

%\cite{tHooft:1976snw}
\bibitem{tHooft:1976snw}
G.~'t Hooft,
%``Computation of the Quantum Effects Due to a Four-Dimensional Pseudoparticle,''
Phys. Rev. D \textbf{14}, 3432-3450 (1976)
[erratum: Phys. Rev. D \textbf{18}, 2199 (1978)]
doi:10.1103/PhysRevD.14.3432
%4612 citations counted in INSPIRE as of 12 May 2025

%\cite{Rehberg:1995kh}
\bibitem{Rehberg:1995kh}
P.~Rehberg, S.~P.~Klevansky and J.~Hufner,
%``Hadronization in the SU(3) Nambu-Jona-Lasinio model,''
Phys. Rev. C \textbf{53}, 410-429 (1996)
doi:10.1103/PhysRevC.53.410
[arXiv:hep-ph/9506436 [hep-ph]].
%399 citations counted in INSPIRE as of 30 Apr 2025

%\cite{Frasca:2011zn}
\bibitem{Frasca:2011zn}
M.~Frasca and M.~Ruggieri,
%``Magnetic Susceptibility of the Quark Condensate and Polarization from Chiral Models,''
Phys. Rev. D \textbf{83}, 094024 (2011)
doi:10.1103/PhysRevD.83.094024
[arXiv:1103.1194 [hep-ph]].
%91 citations counted in INSPIRE as of 28 Apr 2025

%\cite{Ghosh:2019kmf}
\bibitem{Ghosh:2019kmf}
R.~Ghosh, B.~Karmakar and M.~G.~Mustafa,
%``Soft contribution to the damping rate of a hard photon in a weakly magnetized hot medium,''
Phys. Rev. D \textbf{101}, no.5, 056007 (2020)
doi:10.1103/PhysRevD.101.056007
[arXiv:1911.00744 [hep-ph]].
%17 citations counted in INSPIRE as of 17 Apr 2025

%\cite{Ferreira:2014kpa}
\bibitem{Ferreira:2014kpa}
M.~Ferreira, P.~Costa, O.~Louren\c{c}o, T.~Frederico and C.~Provid\^encia,
%``Inverse magnetic catalysis in the (2+1)-flavor Nambu-Jona-Lasinio and Polyakov-Nambu-Jona-Lasinio models,''
Phys. Rev. D \textbf{89}, no.11, 116011 (2014)
doi:10.1103/PhysRevD.89.116011
[arXiv:1404.5577 [hep-ph]].
%216 citations counted in INSPIRE as of 29 Apr 2025

%\cite{Bali:2012zg}
\bibitem{Bali:2012zg}
G.~S.~Bali, F.~Bruckmann, G.~Endrodi, Z.~Fodor, S.~D.~Katz and A.~Schafer,
%``QCD quark condensate in external magnetic fields,''
Phys. Rev. D \textbf{86}, 071502 (2012)
doi:10.1103/PhysRevD.86.071502
[arXiv:1206.4205 [hep-lat]].
%589 citations counted in INSPIRE as of 22 Apr 2025

%\cite{HotQCD:2019xnw}
\bibitem{HotQCD:2019xnw}
H.~T.~Ding \textit{et al.} [HotQCD],
%``Chiral Phase Transition Temperature in ( 2+1 )-Flavor QCD,''
Phys. Rev. Lett. \textbf{123}, no.6, 062002 (2019)
doi:10.1103/PhysRevLett.123.062002
[arXiv:1903.04801 [hep-lat]].
%243 citations counted in INSPIRE as of 30 Apr 2025

%\cite{Ding:2020inp}
\bibitem{Ding:2020inp}
H.~T.~Ding, C.~Schmidt, A.~Tomiya and X.~D.~Wang,
%``Chiral phase structure of three flavor QCD in a background magnetic field,''
Phys. Rev. D \textbf{102}, no.5, 054505 (2020)
doi:10.1103/PhysRevD.102.054505
[arXiv:2006.13422 [hep-lat]].
%31 citations counted in INSPIRE as of 17 Apr 2025

%\cite{Ding:2020hxw}
\bibitem{Ding:2020hxw}
H.~T.~Ding, S.~T.~Li, A.~Tomiya, X.~D.~Wang and Y.~Zhang,
%``Chiral properties of (2+1)-flavor QCD in strong magnetic fields at zero temperature,''
Phys. Rev. D \textbf{104}, no.1, 014505 (2021)
doi:10.1103/PhysRevD.104.014505
[arXiv:2008.00493 [hep-lat]].
%80 citations counted in INSPIRE as of 18 Apr 2025

%\cite{Braun:2020ada}
\bibitem{Braun:2020ada}
J.~Braun, W.~j.~Fu, J.~M.~Pawlowski, F.~Rennecke, D.~Rosenbl\"uh and S.~Yin,
%``Chiral susceptibility in ( 2+1 )-flavor QCD,''
Phys. Rev. D \textbf{102}, no.5, 056010 (2020)
doi:10.1103/PhysRevD.102.056010
[arXiv:2003.13112 [hep-ph]].
%60 citations counted in INSPIRE as of 17 Apr 2025

%\cite{Gao:2021vsf}
\bibitem{Gao:2021vsf}
F.~Gao and J.~M.~Pawlowski,
%``Phase structure of (2+1)-flavor QCD and the magnetic equation of state,''
Phys. Rev. D \textbf{105}, no.9, 094020 (2022)
doi:10.1103/PhysRevD.105.094020
[arXiv:2112.01395 [hep-ph]].
%18 citations counted in INSPIRE as of 09 May 2025

%\cite{Guan:2024ccw}
\bibitem{Guan:2024ccw}
Y.~Guan and S.~Matsuzaki,
%``Ladder top-quark condensation imprints in supercooled electroweak phase transition,''
JHEP \textbf{09}, 140 (2024)
doi:10.1007/JHEP09(2024)140
[arXiv:2405.03265 [hep-ph]].
%4 citations counted in INSPIRE as of 17 Apr 2025

%\cite{Heikinheimo:2018esa}
\bibitem{Heikinheimo:2018esa}
M.~Heikinheimo, K.~Tuominen and K.~Lang\ae{}ble,
%``Hidden strongly interacting massive particles,''
Phys. Rev. D \textbf{97}, no.9, 095040 (2018)
doi:10.1103/PhysRevD.97.095040
[arXiv:1803.07518 [hep-ph]].
%21 citations counted in INSPIRE as of 17 Apr 2025

%\cite{Croon:2019iuh}
\bibitem{Croon:2019iuh}
D.~Croon, R.~Houtz and V.~Sanz,
%``Dynamical Axions and Gravitational Waves,''
JHEP \textbf{07}, 146 (2019)
doi:10.1007/JHEP07(2019)146
[arXiv:1904.10967 [hep-ph]].
%55 citations counted in INSPIRE as of 02 May 2025

%\cite{Pisarski:1983ms}
\bibitem{Pisarski:1983ms}
R.~D.~Pisarski and F.~Wilczek,
%``Remarks on the Chiral Phase Transition in Chromodynamics,''
Phys. Rev. D \textbf{29}, 338-341 (1984)
doi:10.1103/PhysRevD.29.338
%1656 citations counted in INSPIRE as of 30 Apr 2025

%\cite{Kawaguchi:2020kce}
\bibitem{Kawaguchi:2020kce}
M.~Kawaguchi, S.~Matsuzaki and X.~G.~Huang,
%``Dynamic scale anomalous transport in QCD with electromagnetic background,''
JHEP \textbf{10}, 017 (2020)
doi:10.1007/JHEP10(2020)017
[arXiv:2007.00915 [hep-ph]].
%8 citations counted in INSPIRE as of 29 Apr 2025

%\cite{Shifman:1979eb}
\bibitem{Shifman:1979eb}
M.~A.~Shifman, A.~I.~Vainshtein, M.~B.~Voloshin and V.~I.~Zakharov,
%``Low-Energy Theorems for Higgs Boson Couplings to Photons,''
Sov. J. Nucl. Phys. \textbf{30}, 711-716 (1979)
ITEP-42-1979.
%913 citations counted in INSPIRE as of 02 May 2025

%\cite{Matsuzaki:2012vc}
\bibitem{Matsuzaki:2012vc}
S.~Matsuzaki and K.~Yamawaki,
%``Discovering 125 GeV techni-dilaton at LHC,''
Phys. Rev. D \textbf{86}, 035025 (2012)
doi:10.1103/PhysRevD.86.035025
[arXiv:1206.6703 [hep-ph]].
%67 citations counted in INSPIRE as of 09 May 2025

\end{thebibliography}
\end{document}